\documentclass[12pt]{article}

\usepackage[dvips]{graphics,epsfig}
\usepackage[latin1]{inputenc}
\usepackage{setspace}
\usepackage{amsfonts}

\usepackage{textcomp}
\usepackage{amsmath}
\usepackage{graphicx}
\usepackage{braket}
\usepackage{subfig}

\usepackage[unicode=true,bookmarks=false,breaklinks=false,pdfborder={0 0 1},colorlinks=true]
 {hyperref}
\hypersetup{
 citecolor=blue,linkcolor=blue,urlcolor=blue}

\def\laq{\raise 0.4 ex \hbox{$<$}\kern -0.8 em\lower 0.62 ex\hbox{$\sim$}}
\def\gaq{\raise 0.4 ex \hbox{$>$}\kern -0.7 em\lower 0.62 ex\hbox{$\sim$}}

\def\beq{\begin{equation}}
\def\eeq{\end{equation}}
\def\beqa{\begin{eqnarray}} 
\def\eeqa{\end{eqnarray}}
\begin{document}

\pagestyle{plain}

\begin{flushright}
{\bf DRAFT VERSION}\\
April 15, 2021
\end{flushright}
\vspace{0.2cm}

\begin{center}

{\Large\bf Uniformly accelerated quantum counting detector\\
in Minkowski and Fulling vacuum states}

\vspace*{0.5cm}

M. S. Soares$^{*}$, N. F. Svaiter$^{\dagger}$\\
{Centro Brasileiro de Pesquisas F\'{\i}sicas\\
Rua Xavier Sigaud, 150 - Urca, Rio de Janeiro - RJ, 22290-180, Brazil}\\

\vspace*{0.5cm}

C. A. D. Zarro$^{\ddagger}$\\
{ Instituto de F\'isica, Universidade Federal do Rio de Janeiro,\\
Av. Athos da Silveira Ramos 149  Cidade Universit\'aria,\\ Rio de Janeiro - RJ, 21941-909, Brazil
}\\

\vspace*{0.5cm}

G. Menezes$^{\wr}$\\
{Universidade Federal Rural do Rio de Janeiro,\\
 Rodovia BR 465, Km 07,\\
Serop\'edica-RJ, - Brazil}\\
\end{center}

\begin{abstract}
In this work we revisit the discussion on the process of measurements by a detector in a uniformly accelerated rectilinear motion, interacting linearly with a massive scalar field. We employ the Glauber theory of photodetection. In this formalism, the detector, interacting with the field prepared in an arbitrary state of the Rindler Fock space, is excited only by absorption processes. Here we intend to point out the main differences between the Glauber model and the more commonly used Unruh-DeWitt model. To understand more plainly measurement processes within the Glauber framework, we investigate the transition probability rate associated with a uniformly accelerated broadband detector prepared in the ground state. We show how the Unruh-Fulling-Davies effect emerges within the rotating-wave approximation. We also consider the behavior of this broadband detector in a frame which is inertial in the remote past but in the far future becomes uniformly accelerated. The usage of the Glauber model helps to clarify the accelerated or inertial observer interpretations of the measurements performed by the accelerated detector.
 
\end{abstract}



\section{\label{sec:Introduction} Introduction}

In the description of systems with countably infinite number of degrees of freedom, one must define the so-called canonical variables. Then one typically uncovers the algebra determined by such variables and, equipped with such an algebra, one is able to build up the Hilbert space of states of the theory. Nevertheless, it is possible to present unitarily inequivalent representations of the canonical commutation relations \cite{gaardingwightman1954,wightman1955configuration}. The result is that each representation may depict different physical situations. Hence new phenomena such as collective excitations, which can arise in physical systems undergoing spontaneous symmetry breaking, may appear, as is the case of the vacuum of a superconductor with a condensate of Cooper pairs.

It is well known that vacuum fluctuations can inhibit or enhance atomic radiative processes \cite{purcell1946,fordsvaiterlyra1994}. On the other hand, if one evokes the equivalence principle, an important investigation is the study of quantum fields from the perspective of observers in rectilinear uniformly accelerated motion. This leads us to the Fulling quantization \cite{fulling1972,fulling1973} and the Unruh-Davies effect \cite{unruh1976,davies1975,Crispino:2007eb}. According to the Unruh-Davies effect, a detector at rest in a uniformly accelerated frame of reference can be excited by absorption of a Rindler field quantum from the Minkowski vacuum. Our purpose here is to re-examinate the Unruh-Davies effect using the Glauber theory of photodetection \cite{glauber1963A,glauber1963B}. In other words, in this work we revisit the discussion on the process of measuring Rindler field quanta by a uniformly accelerated detector interacting with a massive scalar field within the Glauber model, which realizes detection of quanta as an absorptive process.

The Glauber theory was constructed to formulate models of detectors that respond to the electromagnetic field by absorption of photons. In any case, it is straightforward to implement the same idea to an apparatus device that measures quanta associated with a massive scalar field. In this theory, measurements are absorption processes which are registered, e.g., by the detection of emitted photoelectrons. Such a kind of measurements is known as measurement of second kind, inasmuch as it changes the state being measured. 

We shall address the question of how to define a suitable detector. It is import to bear in mind that measurements of observables in quantum field theory must be performed in some limited space-time domain. From an operational point of view, measurements are means to obtain information about the physical reality using a measurement apparatus. The measurement apparatus sends a sign of information, a signal that must be decoded by some receiver that is able to give us a measurable value at the output of the system. We consider the case in which the dimension of the Hilbert space of the measured system is infinite. In this situation, an ideal detector of field quanta is a dimensionless system that can make a transition between two energy levels, decreasing the number of quanta of the field in some measured state~\cite{moyses1973}. Therefore, a detector of field quanta is an experimental device coupled with the field that gives no signal if the state of the field is the ground state. We should keep in mind that the definition of a vacuum state is related to the choice of time-translation Killing vector fields used by the observers that quantize a classical field.  

The aim of this paper is to explore second-kind measurements in a situation extensively discussed in the literature: Measurements of field quanta associated with a (massive) field performed by observers in rectilinear uniformly accelerated motion. For that matter, it is important to distinguish the Glauber model from the substantially discussed Unruh-DeWitt model, for which absorption and emission processes are equally important. Moreover, since particle detectors are perceived as powerful tools to gather information about quantum fields, it is important to understand the behavior of different models in distinct situations, and we feel that so far the Glauber detector has not received the proper attention in the context of the Unruh-Davies effect. The Glauber model also helps us to understand under an alternative perspective the role of energy transfer between the particle detector and the quantum field in measurement processes. On the other hand, in the quantum theory of photodetection a perturbative treatment of the detector-field interaction using the rotating-wave approximation (RWA) is usually adopted~\cite{glauber1963A,glauber1963B,mandel1964}. In this context, some issues involving causality have already been raised~\cite{bykov1989,tatarskii1990,Martin-Martinez:2015psa,Funai:2019}. However, a thorough analysis reveals that such concerns are unwarranted; indeed, as extensively discussed in Ref.~\cite{milonni1995}, causality is satisfied in the photodetection theory originally formulated by Glauber. This is the interpretation we adopt in this work. For a related discussion of causality in disordered settings, see Ref.~\cite{menezes2017fermi}.

 The Glauber construction introduces a kind of operator ordering where the emission processes are disregarded. We stress that this Glauber ordering is related to the congruence of the timelike Killing vetor field. We evaluate the transition probability rates of the accelerated detector prepared in the ground state for the cases in which the state of the field is taken to be an arbitrary state of $n$-Rindler quanta, a Rindler thermal state with temperature $\beta^{-1}$ and finally the Minkowski vacuum. We recover the well-known result that a detector in a rectilinear uniformly accelerated motion interacting with a field in the Minkowski vacuum has the same transition probability rate of an accelerated detector interacting with the field in a Rindler thermal state at a temperature $\beta^{-1}$. Since there is no agreement in the literature concerning the interpretation of the processes described by observers at rest in an inertial reference frame and at rest in a non-inertial reference frame, using the Glauber model, we intend to shed some light on this issue \cite{unruh1984,ginzburg1987vacuum,barut1990quantum,belyanin2006quantum}. Ultimately, we investigate a model in the Minkowski spacetime where the vacuum defined in the remote past, the state $\ket{0, \text{in}}$, and in the far future, the state $\ket{0, \text{out}}$, are unitarily non-equivalent. We show that a Glauber detector in the far future can be excited when interacting with the field in the vacuum state $\ket{0, \text{in}}$. 

The organization of this paper is as follows. In Sec. \ref{sec:fullingquantization}, the Fulling quantization, suitable for uniformly accelerated observers and for addressing the Unruh-Davies effect, is briefly reviewed. In Sec. \ref{sec:glauberdetector} the Glauber theory of detection of field quanta, interpreting measurements by absorption of field quanta, is presented. In Sec. \ref{sec:glauberdetectorunruh} we perform the analysis of the radiative processes of the detector in a uniformly accelerated frame. In Sec. \ref{sec:kalnis}, we discuss a Glauber detector for a Kalnins-Miller coordinate system.  In Sec. \ref{sec:conclusions} we present our conclusions. We use units such that $\hbar = c = k_{B} = 1$ and $\eta_{\mu \nu} = diag(+1, -1,-1,-1)$.


\section{The Fulling quantization and the Unruh-Davies effect}\label{sec:fullingquantization}

In this section, we briefly review the discussion on the Unruh-Davies effect, with the Fulling quantization of a scalar field in Rindler spacetime \cite{rindler1966}. See also Refs. \cite{dewitt1975,sciamacandelasdeutsch1981,takagi1986}. In order to make the debate accessible to a general readership not very familiar with some technical jargon usually employed in this context, we will give here a quick, self-contained examination of this topic, even though this is a somewhat standard discussion. In this way we believe our conclusions will be expounded in a clear way. The main result discusses the behavior of an Unruh-DeWitt detector at rest in a frame of reference with a rectilinear uniformly accelerated motion. As well known, uniformly accelerated observers in Minkowski spacetime envisage the Minkowski vacuum as a thermal bath of Rindler particles. 

Since in an arbitrary globally hyperbolic stationary spacetime one can always find a Killing vector field corresponding to the time direction of some family of inertial observers, in Minkowski spacetime one can use this feature to decompose the free scalar field operator into a sum of its positive and negative frequency parts:
\begin{equation}
    \varphi(t,\mathbf{x})=\varphi^{(+)}(t,\mathbf{x})+\varphi^{(-)}(t,\mathbf{x})
\end{equation}
where
\begin{equation}\label{eq:phimink+}
    \varphi^{(+)}(t,\mathbf{x})=\frac{1}{(2 \pi)^{3/2}}\int\frac{d^{3}k}{\sqrt{2\omega_{k}}}a(\mathbf{k})e^{- i(\omega_{k}t -\mathbf{k}\cdot\mathbf{x})}
\end{equation}
and
\begin{equation}\label{eq:phimink-}
    \varphi^{(-)}(t,\mathbf{x})=\frac{1}{(2 \pi)^{3/2}}\int\frac{d^{3}k}{\sqrt{2\omega_{k}}}a^{\dagger}(\mathbf{k})e^{i(\omega_{k}t -\mathbf{k}\cdot\mathbf{x})}
\end{equation}
for $\omega_{k}=\sqrt{\mathbf{k}^{2}+m_{0}^{2}}$, $m_0$ being the mass of the field. These modes are normalized using the standard inner product \cite{Bire82}. The annihilation and creation operators for quanta of the field, $a(\mathbf{k})$ and $a^{\dagger}(\mathbf{k})$ satisfy the usual commutation relations.
The vacuum state built by observers at rest in an inertial reference frame that have a timelike Killing vector field is the state in which field quanta are absent. In conclusion, there is a translational-invariant vacuum state $\ket{0,M}$ such that $a(\mathbf{k}) \ket{0,M } = 0\ \ \forall \mathbf{k}$. 

Consider now a family of observers at rest in a non-inertial reference frame, $e.g.$, with a rectilinear uniformly accelerated motion. Starting from the usual Cartesian coordinates $x^{\mu} = (t,x^1, x^2, x^3)$, one defines the curvilinear coordinates $X^{\mu} = (\eta, \xi, y, z)$ using
\begin{equation}
    \begin{cases} t=\xi \sinh{\eta} \\ x^{1}=\xi \cosh{\eta}  \\ x^{2}=y \\ x^{3}=z,\end{cases}
\end{equation}
where $0<\xi<\infty$ and $- \infty < \eta < \infty$. Therefore $\xi^2= (x^{1})^{2}-t^{2}$ and $\eta=\tanh^{-1}(t/(x^{1}))$. An observer travelling in the wordline $\xi=1/a = \text{constant}$ ($y,z$ are also constants) has proper acceleration given by $a$. This coordinate system covers only a wedge of the Minkowski spacetime, $i.e.$ the region $|x|>t$, where there is a global timelike Killing vector field $\partial/\partial \eta$. 

We restrict our considerations to the right Rindler wedge where the detector is traveling. Define the vectors $\mathbf{q}=(k_{y},k_{z})$ and $\mathbf{y}=(y,z)$.  From the general solutions of the Klein-Gordon equations, the scalar field operator can be written as sum of positive and negative frequency contribution with respect to $\partial/\partial \eta$ as $\varphi(\eta,\xi,\mathbf{y}) = \varphi^{(+)}(\eta,\xi,\mathbf{y}) + \varphi^{(-)}(\eta,\xi,\mathbf{y})$. Defining $m^2 = m_{0}^{2}+\mathbf{q}^{2}$, we have
\begin{align}
    \varphi^{(+)}(\eta,\xi,\mathbf{y})&=\frac{1}{2\pi^2}\int_{0}^{\infty}d\nu\int d^{2}\mathbf{q} \sqrt{\sinh{\pi\nu}}K_{i\nu}(m\xi)\nonumber \\
    &e^{-i(\nu\eta-\mathbf{q}\cdot\mathbf{y})} b(\nu,\mathbf{q}),\label{eq:phirindler+}
\end{align}
and
\begin{align}
      \varphi^{(-)}(\eta,\xi,\mathbf{y})&=\frac{1}{2\pi^2}\int_{0}^{\infty}d\nu\int d^{2}\mathbf{q} \sqrt{\sinh{\pi\nu}}K_{i\nu}(m\xi)\nonumber \\
    &e^{i(\nu\eta-\mathbf{q}\cdot\mathbf{y})} b^{\dagger}(\nu,\mathbf{q}), \label{eq:phirindler-}
\end{align}
where $K_{i \nu}(z)$ is the Macdonald function of imaginary order. The annihilation and creation operator of Rindler field quanta $b(\nu,\mathbf{q})$ and $b^{\dagger}(\nu,\mathbf{q})$ satisfy the usual commutation relations over a constant $\eta$ hypersurface. It is clear that in the Hilbert space of states there is a vacuum state $|0,R\rangle$, known as the Fulling vacuum state, such that
$b(\nu,\mathbf{q})|0,R\rangle=0,\quad \forall \nu \in [0,\infty), -\infty<k_{y},k_{z}<\infty$. From the Fulling vacuum one can generate the Rindler Fock space.  

The annihilation operator of Rindler field quanta in the mode $(\nu, \mathbf{q})$ can be expanded, using Bogoliubov transformations, into a linear combination of creation and annihilation of Minkowski field operators $a^{\dagger}(\mathbf{k})$ and $a(\mathbf{k})$. We have
\begin{equation}
    b(\nu,\mathbf{q})=\int d^{3}\mathbf{k}\left[U(\nu,\mathbf{q},\mathbf{k})a(\mathbf{k})+V(\nu,\mathbf{q},\mathbf{k})a^{\dagger}(\mathbf{k})\right] \label{eq:bogoliubov1},
\end{equation}
where
\begin{align}
    U(\nu,\mathbf{q},\mathbf{k}')&=\frac{1}{\left[2\pi\omega_{k}(1-e^{-2\pi\nu})\right]^{\frac{1}{2}}}\left(\frac{\omega_{k}+\vert \mathbf{k}\vert}{m}\right)^{i\nu}\delta(\mathbf{q}-\mathbf{q}'), \label{eq:bogoliubovalpha}\\
    V(\nu,\mathbf{q},\mathbf{k}')&=\frac{1}{\left[2\pi\omega_{k}(e^{2\pi\nu}-1)\right]^{\frac{1}{2}}}\left(\frac{\omega_{k}+\vert \mathbf{k}\vert}{m}\right)^{i\nu}\delta(\mathbf{q}-\mathbf{q}') \label{eq:bogoliubovbeta}.
\end{align}
So we see that, in the construction of the mathematical formalism, we obtain two inequivalent Fock spaces with two vacuum states $\ket{0,M}$ and $\ket{0,R}$. It turns out that the Poincar\'{e} invariant vacuum $\ket{0,M}$ has an infinite number of Rindler quanta in the mode $(\nu, \mathbf{q})$ associated with the massive scalar field:
\begin{align}\label{eqq:bedistribution}
    \bra{0,M}b^{\dagger}(\nu,\mathbf{q}) &b(\nu^{\prime},\mathbf{q}^{\prime})\ket{0,M}\nonumber \\ &= \frac{1}{e^{2\pi\nu}-1}\delta\left(\nu - \nu^{\prime}\right) \delta\left(\mathbf{q} - \mathbf{q}^{\prime}\right).
\end{align}

We now give a crash course of basic results concerning a two-level Unruh-DeWitt detector in the light of time-dependent perturbation theory. It consists of an idealized point-like two-level system with a ground state $\ket{g}$, with energy $\omega_{g}$, and an excited one $\ket{e}$, with energy $\omega_{e}$. The detector-field interaction Hamiltonian is given by
\begin{equation}\label{eq:hintdetc-field}
    H_{\text{int}}=c_{1}m(\tau)\varphi(x(\tau)),
\end{equation}
where $m(\tau)$ is called the monopole operator and $c_1$ is a dimensionless small coupling constant. Here $\tau$ is the proper time of the detector. The total Hamiltonian of the system can be written as $H = H_{F} + H_{D} + H_{\text{int}}$, where it is also composed by the noninteracting detector Hamiltonian $H_{D}$ and the free scalar field Hamiltonian $H_{F}$.

Let us assume here the adiabatic hypothesis. We are interested to discuss the behavior of the detector interacting with the field in prepared states. The temporal dependence of $H_{\text{int}}$ is given by 
    \begin{equation}\label{eqq:evolutionofhint}
        H_{\mathrm{int}}(\tau)=e^{i H_{0} \tau}\left(H_{\mathrm{int}}\right)_{S} e^{-i H_{0} \tau},
    \end{equation}
where $\left(H_{\mathrm{int}}\right)_{S}$ is the interaction Hamiltonian in the Schrödinger picture and $H_{0}$ is the nonperturbate Hamiltonian of the system, $H_{0} = H_{F} + H_{D}$. The system obeys the equations
    \begin{eqnarray}
    i\frac{\partial}{\partial \tau} \ket{\tau} = H_{\text{int}} \ket{\tau},\\
    \ket{\tau} = U(\tau, \tau_{i})\ket{\tau_{i}},
    \end{eqnarray}
with $U(\tau,\tau_{i})$ being the evolution operator. In first-order approximation, the evolution operator becomes
    \begin{equation}\label{eqq:evolutionoperator1ord}
        U(\tau_{f}, \tau_{i})=1-i \int_{\tau_{i}}^{\tau_{f}} d \tau^{\prime} H_{\mathrm{int}}\left(\tau^{\prime}\right).
    \end{equation}
By using the following operators
\begin{eqnarray}
\sigma_{z} &=& \frac{1}{2} \Bigl( | e \rangle \langle e | 
- | g \rangle \langle g | \Bigr)
\nonumber\\
\sigma^{+} &=& | e \rangle \langle g |
\nonumber\\
\sigma^{-} &=& | g \rangle \langle e |
\label{sigma}
\end{eqnarray}
which satisfy known angular momentum commutation relations, the detector-field interaction Hamiltonian can be rewritten as
\begin{equation}\label{eq:hintdetc-field2}
    H_{\text{int}}=c_{1} \bigl[ m_{eg} \sigma^{+} 
    + m_{ge} \sigma^{-} + \sigma_{z} (m_{ee} - m_{gg}) \bigr] \varphi(x(\tau)),
\end{equation}
where $m_{ij}=\bra{i}m(0)\ket{j}$, $|i \rangle = | e,g \rangle$. When we perform the aforementioned field decomposition into a sum of positive and negative frequency pieces in Eq.~(\ref{eq:hintdetc-field2}), we find that the leading-order contributions to the field-detector state display four kinds of terms associated with absorption or emission process of quanta of the field with the excitation or decay of the detector~\cite{svaiter1992}. The Unruh-DeWitt detection model takes into account all such contributions. 

Using Eq. \eqref{eqq:evolutionoperator1ord}, one can find the transition probability from the initial state $\ket{\tau_{i}} = \ket{g}\otimes\ket{\Phi_{i}}$ at $\tau_{i}$ to a final state $\ket{\tau_{f}} = \ket{e}\otimes\ket{\Phi_{f}}$ at $\tau_{f}$, where $|\Phi_{i}\rangle$ is the initial field state and $\ket{\Phi_{f}}$ is the final field state by computing the amplitude probability transition given by
    \begin{align}
       \bra{\tau_{f}} U(\tau_{f}, \tau_{i})\ket{\tau_{i}} = -i c_{1} \int_{\tau_{i}}^{\tau_{f}}\bra{\tau_{f}} m\left(\tau^{\prime}\right) \phi\left(x\left(\tau^{\prime}\right)\right)\ket{\tau_{i}} d \tau^{\prime}. 
    \end{align}
Therefore, using Eq. \eqref{eqq:evolutionofhint}, the amplitude probability of transitions becomes
    \begin{align}
        &A_{\ket{\tau_{i}}\rightarrow\ket{\tau_{f}}} \nonumber\\
        &= -i c_{1} \int_{\tau_{i}}^{\tau_{f}}d \tau^{\prime}~ e^{i \omega_{eg} \tau^{\prime}}\bra{e} m(0)\ket{g}\bra{\Phi_{f}}\varphi\left(\tau^{\prime}, \mathbf{x}\right)\ket{\Phi_{i}}, \label{eqq:amplitudeoftransitions}
    \end{align}
where $\omega_{eg} = \omega_{e} - \omega_{g}$ is the energy gap between the energy levels of the detector. Therefore, by Eq. \eqref{eqq:amplitudeoftransitions}, and after summing over a complete set $\{\ket{\Phi_{f}}\}$, one obtains the probability of transitions as
    \begin{eqnarray}
    \hspace{-6mm}
        P_{\ket{\tau_{i}}\rightarrow\ket{\tau_{f}}}(\tau_{f}, \tau_{i})&=& \left| A_{\ket{\tau_{i}}\rightarrow\ket{\tau_{f}}}\right|^{2}
        \nonumber\\
&=& c_{1}^{2}\left|\bra{e}m(0)\ket{g}\right|^{2}F(\omega_{eg},\tau_{f},\tau_{i},\mathbf{x}) ,
        \end{eqnarray}
where  $c_{1}^{2}\left|\bra{e}m(0)\ket{g}\right|^{2}$ is called selectivity of the detector and $F(\omega_{eg},\tau_{f},\tau_{i},\mathbf{x})$ is the so-called response function which is, in first-order perturbation theory:
\begin{align}
    F(\omega_{eg},\tau_{f},\tau_{i},\mathbf{x})&=\int_{\tau_{i}}^{\tau_{f}}d\tau\int_{\tau_{i}}^{\tau_{f}}d\tau' e^{-i\omega_{eg}(\tau-\tau')}\nonumber \\
    &\langle \Phi_{i}|\varphi(\tau,\mathbf{x})\varphi(\tau',\mathbf{x})|\Phi_{i}\rangle,
\end{align}
where we used that $\sum_{f}|\Phi_{f}\rangle \langle \Phi_{f}|=1$. The response function reveals the bath of quanta that the detector may experience. Since measurements in field theory must be evaluated in a limited time interval, for the case of a uniformly accelerated Unruh-DeWitt detector and a real massless scalar field in the Minkowski vacuum, the transition probability was evaluated for a finite proper time in Refs. \cite{svaiter1992,Sriramkumar:1994pb}. The asymptotic rate of spontaneous and induced emission and absorption of Rindler field quanta (transition probability per unit proper time in the limit $\tau_{f} - \tau_{i} \to \infty$) is given by  
\begin{align}
R(\omega_{eg})&=\frac{|\omega_{eg}|}{2\pi}\left[\theta(-\omega_{eg})\left(1+\frac{1}{e^{2\pi \sigma \omega_{eg}}-1}\right)\right. \nonumber \\
&+\left. \theta(\omega_{eg})\frac{1}{e^{2\pi \sigma \omega_{eg}}-1}\right],
\end{align}
where $\sigma^{-1}$ is the proper acceleration of the detector. The common lore in the literature on the Unruh effect states that the first term should describe spontaneous emission of the uniformly accelerated detector, whereas the second term represents spontaneous excitation. For related discussions, see also Refs.~\cite{Audretsch:94,Shih-Yuin:06,Shih-Yuin:07,Zhu:07,Hu:12,Menezes:2015iva, Menezes:2015uaa,Fewster:16,Carballo-Rubio:19}. The interpretation of how the process of measurement is described by different observers is far from being obvious. Hence a fundamental question would be how one should interpret detector measurements in different frames. This is the topic of the next sections.


\section{The Glauber theory of photodetection}\label{sec:glauberdetector}

In this section we are interested to discuss measurements in the scenario of radiative processes of prepared states. The purpose here is to emphasize several features of the Glauber theory of photodetection that will be important in the analysis of the Unruh-Davies effect within this formalism. Some of the points discussed here are not recurrently found in the associated literature concerning this subject. We wish to consider the case of real atoms for which, e.g., the continuum of excited states (in a single electron atom) corresponds to electronic states above the ionization threshold.

The detector Hilbert space of states is the ground state $|\psi_{b}\rangle$ and a continuum of unbounded states $|\psi_{j}\rangle$. The time independent detector Hamiltonian satisfies
\begin{equation}
    H_{d}|\psi_{l}\rangle=E_{l}|\psi_{l}\rangle,\ l = b \text{\ or\ }j
\end{equation}
where $E_b$ ($E_j$) is the energy associated with the ground state (excited states). The detector-field interaction Hamiltonian is still given by Eq. \eqref{eq:hintdetc-field} and the monopole operator now reads
\begin{equation}
m=\int dj \bigg( m_{jb}\ket{\psi_{j}}\bra{\psi_{b}}+m_{bj}\ket{\psi_{b}}\bra{\psi_{j}}\bigg),    
\end{equation}
where $m_{ik}=\bra{\psi_{i}}m(\tau_{i})\ket{\psi_{k}}$. Notice that  $\tau$ is the detector's proper time. If we are not using $\tau$ as the detector proper time, an appropriate factor will need to be introduced in the interaction Hamiltonian \cite{Martin-Martinez:2018gzb}. 

The key distinction between the Glauber and the Unruh-DeWitt detection models can be seen by considering an alternative form for the detector-field interaction Hamiltonian similar to the one given by Eq.~(\ref{eq:hintdetc-field2}). In the present case, this reads
\begin{equation}\label{eq:hintdetc-field3}
    H_{\text{int}}=c_{1} \int dj \bigl[ m_{jb} \sigma^{+} 
    + m_{bj} \sigma^{-} + \sigma_{z} (m_{jj} - m_{bb}) \bigr] \varphi(x(\tau))
\end{equation}
where the operators $\sigma_{z}, \sigma^{\pm}$ are now written in terms of $|\psi_{j}\rangle$ and $|\psi_{b}\rangle$ but maintain the same form as the one given in Eq.~(\ref{sigma}). If again we decompose the field into a sum of positive and negative frequency parts, once more we find that the leading-order contributions to the field-detector state unveil four kinds of terms which have the same interpretation as above. Two of such terms will have the general form $c^{\dagger}({\bf k}) \sigma^{+}$ and $c({\bf k}) \sigma^{-}$, where $c^{\dagger}, c$ stand for generic creation and annihilation field operators (could be either Minkowski operators or Rindler operators, depending on the context). These are known as antiresonant terms and are completely disregarded by the rotating-wave approximation. Hence, while the Unruh-DeWitt detector maintains all such terms, as mentioned above, the Glauber detector only considers the so-called resonant terms, $c^{\dagger}({\bf k}) \sigma^{-}$ and $c({\bf k}) \sigma^{+}$.

One can find the transition amplitude from the initial state $\ket{\tau_{i}} = \ket{\psi_{b}}\otimes\ket{\Phi_{i}}$ at $\tau_{i}$ to $\ket{\tau_{f}} = \ket{\psi_{j}}\otimes\ket{\Phi_{f}}$ at $\tau_{f}$, where $|\Phi_{i}\rangle$ and $|\Phi_{f}\rangle$ are arbitrary pure field states. For instance, they can be Fock states constructed with $a^{\dagger}({\mathbf{k}})$ operators acting in the Poincar\'{e} invariant vacuum state $\ket{0,M}$. Using a similar procedure as above, one finds that (at first order)
    \begin{align}
        &A_{\ket{\tau_{i}}\rightarrow\ket{\tau_{f}}} \nonumber\\
        &= -i c_{1} \int_{\tau_{i}}^{\tau_{f}}d \tau^{\prime}~ e^{i \omega_{jb} \tau^{\prime}}\bra{\psi_{j}} m(0)\ket{\psi_{b}}\bra{\Phi_{f}}\varphi\left(\tau^{\prime}, \mathbf{x}\right)\ket{\Phi_{i}}, \label{eq:amplitudeoftransitions}
    \end{align}
where the energy gap is now defined by $\omega_{jb} = E_{j} - E_{b}$. As discussed in Ref.~\cite{svaiter1992}, after distinguishing positive and negative frequency contributions coming from the field operator, one verifies the presence of two terms, namely
    \begin{align}
\bra{\Phi_{f}}\varphi^{+}\left(\tau^{\prime}, \mathbf{x}\right)\ket{\Phi_{i}}
+ 
\bra{\Phi_{f}}\varphi^{-}\left(\tau^{\prime}, \mathbf{x}\right)\ket{\Phi_{i}}
\label{pm}
    \end{align}
which are associated, respectively, with the absorption and emission of inertial quanta of the field with excitation of the detector. Notice that, when considering photodetection as the photoabsorption process, in the probability amplitude only the first term above is relevant. In addition, observe that, for excitation processes, $\omega_{jb} > 0$. It is the case we are interested in. In any case, we notice that such processes are defined using a particular timelike Killing vector field and the positive frequency modes are associated with this timelike Killing vector field.  

In order to highlight even more the differences between the behavior of the two forementioned detector models in radiative processes, one can also investigate the associated response function. As above, this is related to the
transition probability which is 
$P_{\ket{\tau_{i}}\rightarrow\ket{\tau_{f}}}(\tau_{f}, \tau_{i})=\left| A_{\ket{\tau_{i}}\rightarrow\ket{\tau_{f}}}\right|^{2}$. After summing over the complete set of final field states 
$\{ |\Phi_{f}\rangle \}$, we get
    \begin{align}
         P_{\ket{\tau_{i}}\rightarrow\ket{\tau_{f}}}(\tau_{f}, \tau_{i})=c_{1}^{2}\left|\bra{\psi_{j}}m(0)\ket{\psi_{b}}\right|^{2}F(\omega_{eg},\tau_{f},\tau_{i},\mathbf{x}),
    \end{align}
where the response function now reads
\begin{align}
    F(\omega_{jb},\tau_{f},\tau_{i},\mathbf{x})&=\int_{\tau_{i}}^{\tau_{f}}d\tau^{\prime}\int_{\tau_{i}}^{\tau_{f}}d\tau^{\prime \prime} e^{-i\omega_{jb}(\tau^{\prime}-\tau^{\prime \prime})}\nonumber \\
    & \times \langle \Phi_{i}|\varphi(\tau^{\prime},\mathbf{x})\varphi(\tau^{\prime \prime},\mathbf{x})|\Phi_{i}\rangle.
\end{align}
Using again the decomposition of the field operator into the sum of positive and negative frequency parts, the response function can be written as
\begin{equation}
    F(\omega_{jb},\tau_{f},\tau_{i},\mathbf{x}) = F_{1}(\omega_{jb},\tau_{f},\tau_{i},\mathbf{x})+F_{2}(\omega_{jb},\tau_{f},\tau_{i},\mathbf{x}),
\end{equation}
where
\begin{align}
    F_{1}(\omega_{jb},\tau_{f},\tau_{i},\mathbf{x})&=\int_{\tau_{i}}^{\tau_{f}}d\tau^{\prime}\int_{\tau_{i}}^{\tau_{f}}d\tau^{\prime \prime} e^{-i\omega_{jb}(\tau^{\prime}-\tau^{\prime \prime})}\nonumber \\
                            & \times \langle \Phi_{i}|\varphi^{(-)}(\tau^{\prime},\mathbf{x})\varphi^{(+)}(\tau^{\prime \prime},\mathbf{x})|\Phi_{i}\rangle,
\end{align}
and
\begin{align}
     F_{2}(\omega_{jb},\tau_{f},\tau_{i},\mathbf{x})&=\int_{\tau_{i}}^{\tau_{f}}d\tau^{\prime}\int_{\tau_{i}}^{\tau_{f}}d\tau^{\prime \prime} e^{-i\omega_{jb}(\tau^{\prime}-\tau^{\prime \prime})}\nonumber \\
                            & \times \langle \Phi_{i}|\varphi^{(+)}(\tau^{\prime},\mathbf{x})\varphi^{(-)}(\tau^{\prime \prime },\mathbf{x})|\Phi_{i}\rangle.
\end{align}
Our definition of a \textit{bona fide} detector is a device that goes to an excited state by decreasing the number of quanta of some state. Since $F_{1}(\omega_{jb},\tau_{f},\tau_{i},\mathbf{x})$ is constructed solely from the first term of Eq.~(\ref{pm}), it is clear that it only describes absorption processes. In the Glauber theory of photodetection only this term contributes to the transition rate. 

On the other hand,
$F_{2}(\omega_{jb},\tau_{f},\tau_{i},\mathbf{x})$ is obtained from the second term of Eq.~(\ref{pm}), and as a consequence it is commonly associated with emission processes, accompanied by the decay of the detector. However, since we are considering $\omega_{jb} > 0$, one may ask whether one should really maintain such a contribution. Indeed, as shown in Ref. \cite{svaiter1992}, the $F_{2}$ contribution vanishes in the asymptotic limits $\tau_i \to -\infty, \tau_f \to \infty$ when $\varphi^{(+)}(x)$ and $\varphi^{(-)}(x)$ are given by Eqs.~(\ref{eq:phimink+}) and~(\ref{eq:phimink-}) and $|\Phi_{i}\rangle$ is the Minkowski vacuum state. Furthermore, there are additional terms when using the Unruh-DeWitt detector that vanish only in the asymptotic limits $\tau_i \to -\infty, \tau_f \to \infty$~\cite{svaiter1992}. Such terms are not connected with absorption of field quanta.

 An idealized broadband detector occurs when many final states with range of energies much wider than the bandwidth of the radiation field contribute to the absorption processes \cite{moyses1973,knight1983}. See also Refs. \cite{Montero:2012yv,Ralph:2015vra,Roussel:2019ban}. Moreover, the energy range is much wider than reciprocal of the detection time. Since this is the common situation in detection processes, let us assume a broadband detector. For simplicity we choose $\tau_{i} = 0$ and $\tau_{f} = \tau$. The density of final excited states of the detector is defined by $\rho(\omega_{jb})$ \cite{knight1983}. The probability of excitation is
\begin{equation}
    P(\tau, \mathbf{x})=\int\rho(\omega_{jb}) P_{|\psi_{b}\rangle\to|\psi_{j}\rangle}(\tau,\mathbf{x})\,d\omega_{jb}.
\end{equation}
Considering that $\rho(\omega_{jb})$ is a slowly-varying function, it can be replaced by a constant value and one finds
\begin{equation}
   \int_{-\infty}^{\infty} d\omega_{jb} e^{i\omega_{jb}(\tau''-\tau')}\rho(\omega_{jb})=2\pi\delta(\tau''-\tau')\rho(\bar{\omega}_{jb}).
\end{equation}
In conclusion, for a broadband detector, with only the
$F_{1}(\omega_{jb},\tau_{f},\tau_{i},\mathbf{x})$ contribution, one can show that the excitation probability is given by
\begin{equation}
    P(\tau, \mathbf{x})=C\int_{0}^{\tau}d\tau'\langle \Phi_{i}|\varphi^{(-)}(\tau',\mathbf{x})\varphi^{(+)}(\tau',\mathbf{x})|\Phi_{i}\rangle,
\end{equation} 
where $C$ represents the efficiency of the detector.  The probability of excitation per unit time at time $\tau$ is defined by $d P(\tau,\mathbf{x}) / d\tau$. Therefore, the transition probability rate of the detector reads
\begin{align}
    \frac{dP}{d\tau} = W(\tau, \mathbf{x}; \ket{\Phi_{i}})&=C\bra{\Phi_{i}}\varphi^{(-)}(\tau,\mathbf{x})\varphi^{(+)}(\tau,\mathbf{x})\ket{\Phi_{i}}.
\end{align}
This is the standard result of the Glauber quantum counting model. The transition probability rate comes from absorption processes, where $\varphi^{(+)}(\tau, \mathbf{x})$ and $\varphi^{(-)}(\tau, \mathbf{x})$ are given by Eqs. \eqref{eq:phimink+} and 
\eqref{eq:phimink-}.

We close this section with a brief discussion on possible causality violation in the Glauber's detector model. As discussed in the literature, when antiresonant terms are dropped from the interaction Hamiltonian, the field operators evaluated from this approximate Hamiltonian are not retarded~\cite{Compagno:90}. This will imply in causality violations and possible faster-than-light signaling between two detectors interacting with a common quantum field. This is precisely what is found in Refs.~\cite{Martin-Martinez:2015psa,Funai:2019}. However, if one employs an alternative description for the rotating-wave approximation -- that is, to consider the neglect of counter-rotating terms only after calculating transition amplitudes -- then a proof of causality can be given by using the standard Fermi-problem set up~\cite{milonni1995}. This can be easily verified in the situation in which one resorts to a precise specification of the state of the detectors and also of the quantum scalar field at the initial time $\tau_0$ as well as at a later time $\tau$. The more general case in which only the state of one of the detectors is specified at time $\tau$ is more involved and should be analyzed carefully. This would be interesting to explore, and we hope to return to this calculation in the future.


\section{The Glauber theory of photodetection in the Unruh-Davies effect}\label{sec:glauberdetectorunruh}

We are now in the position to investigate radiative processes of the Glauber detector at rest in a uniformly accelerated frame of reference. In the Glauber theory the detector works only by absorption of field quanta. In order to understand the essential point of our approach recall that in, stationary spacetimes such as the Minkowski or Rindler spacetime, observers use different choices of timelike Killing fields. Therefore one must define two different orderings, given by $:\varphi^{(-)}(x)\varphi^{(+)}(x'):{\text{\tiny{M}}}$, using Eqs. \eqref{eq:phimink+} and \eqref{eq:phimink-} or $:\varphi^{(-)}(x)\varphi^{(+)}(x'):{\text{\tiny{R}}}$, using Eqs. \eqref{eq:phirindler+} and \eqref{eq:phirindler-} related to each Killing vector field respectively.

Let us study the broadband detector introduced above, but now we explicitly take into account a uniformly accelerated motion. In this case the transition probability rate must be given by products of $\varphi^{(+)}(x)$ and $\varphi^{(-)}(x)$ defined in Eqs.~\eqref{eq:phirindler+} and~(\ref{eq:phirindler-}), respectively. First we will consider that the field is in an arbitrary state $|\phi_{i}\rangle$ constructed with the $b^{\dagger}(\nu)$ operators acting on the Fulling vacuum state $|0,R\rangle$. Since we are using the above defined ordering associated with the Rindler timelike Killing vector field, similar steps as those previously outlined lead us to the following expression for the transition probability rate $ W(\eta, \xi; \ket{\phi_{i}})$:
\begin{align}\label{eq:Wfinal}
    W(\eta, \xi; \ket{\phi_{i}})&=\bra{\phi_{i}}\varphi^{(-)}(\eta,\xi,\mathbf{y})\varphi^{(+)}(\eta,\xi,\mathbf{y})\ket{\phi_{i}}
\end{align}
where we are employing the Rindler time $\eta$ in order to describe the time evolution of the system and for simplicity we have set $C=1$. Observe that the average Rindler quantum counting rate is proportional to the expectation value of the normal ordered product of the negative and positive frequency parts contributions of the field operator at the worldline $\xi = \text{constant}$ and $y,z$ also constant. Using Eqs.~(\ref{eq:phirindler+}) and~(\ref{eq:phirindler-}), one finds that
\begin{align}
W(\xi; \ket{\phi_{i}}) &= 
    \frac{1}{4\pi^4}\int_{0}^{\infty}d\nu
    \int_{0}^{\infty}d\nu^{\prime}
    \int d^{2}\mathbf{q} 
    \int d^{2}\mathbf{q}^{\prime}
\nonumber \\    
    & \times e^{i[\eta(\nu-\nu^{\prime}) 
    - \mathbf{y} \cdot (\mathbf{q}-\mathbf{q}^{\prime}) ]} \sqrt{\sinh{\pi\nu}}\sqrt{\sinh{\pi\nu^{\prime}}}
\nonumber \\
    &\times  K_{i\nu}(m\xi) K_{i\nu^{\prime}}(m\xi) \bra{\phi_{i}}b^{\dagger}(\nu, \mathbf{q})b(\nu^{\prime},\mathbf{q}^{\prime})\ket{\phi_{i}}.
\label{transitionrate}
\end{align}
It is now clear that, when the state of the Rindler Fock space is the Fulling vacuum, $i.e.$, $|\phi_{i}\rangle=|0,R\rangle$, the rate of excitation vanishes, as expected.

Assume now a thermal bath in Rindler spacetime. Each observer at rest in an accelerated frame of reference measures a local temperature $T = \beta^{-1}$. Planck's law shows that the transition rate in a specific worldline is
\begin{align}\label{eq:wbeta}
 W_{\beta}(\xi) = \frac{1}{4\pi^{4}}\int d^{2}\mathbf{q}\int_{0}^{\infty}d\nu\frac{\sinh{\pi\nu}}{e^{\beta\nu}-1}K_{i\nu}^{2}\left(\xi\sqrt{m_{0}^2 + \mathbf{q}^2}\right).
\end{align}
Let us make the connection between temperature and proper acceleration. A family of observers can be defined by a set of timelike worldlines. In Rindler spacetime each observer travelling in the worldline $\xi = \text{constant}$ defines a uniformly accelerated frame with proper acceleration $a = \xi^{-1}$. One can show that there is a thermal equilibrium using the Tolman relation $\beta^{-1}\sqrt{g_{00}} = \text{constant}$ \cite{PhysRev.35.875, PhysRev.36.1791}. Therefore, the local temperature in each wordline is $\beta^{-1} = (2 \pi \xi)^{-1}$.

Now we want to understand how the detector behaves if one substitutes the Rindler Fock space by another Hilbert space that carry another representation of the field algebra. This is the fundamental point here. One is applying operators constructed to act in one representation of the operator algebra to states that belong to an unitarily inequivalent representation. In other words, we wish to comprehend how the uniformly accelerated Glauber detector behaves if the state of field is the Minkowski vacuum. Starting from Eq.~(\ref{transitionrate}) using that $|\phi_{i}\rangle=|0,M\rangle$ we obtain the transition rate of the uniformly accelerated dectetor interacting with the scalar field in the Minkowski vacuum. Let us call it $W_{1}( \xi; \ket{0,M})$. Now using Eq. \eqref{eqq:bedistribution}, one obtains
\begin{align}\label{eq:excitationratenonrenormalized}
    W_{1}( \xi; \ket{0,M})
    &=  \frac{1}{4\pi^{4}}\int d^{2}\mathbf{q}\int_{0}^{\infty}d\nu\frac{\sinh{\pi\nu}}{e^{2\pi\nu}-1}\nonumber \\ & \times K_{i\nu}^{2}\left(\xi\sqrt{m_{0}^2 + \mathbf{q}^2}\right).
\end{align}
We obtain that $ W_{1}( \xi; \ket{0,M})= W_{\beta}( \xi)$. This is another version of the Bisognano-Wichmann theorem \cite{Bisognano:1975ih,Bisognano:1976za} that states that the Minkowski vacuum expectation value of observables which are localized in the right Rindler wedge satisfies the Kubo-Martin-Schwinger condition \cite{Kubo:1957mj,Martin:1959jp} with respect to the Rindler time variable $\eta$. From the point of view of the accelerated observer, the contribution to the rate given by $W_{1}$ is a process of excitation of the Glauber detector with absorption of a Rindler quantum from the scalar field. 

From Eq. \eqref{eq:excitationratenonrenormalized}, for the case where $m_0^{2} = 0$ we obtain the well-known result
\begin{align}\label{eq:masslesslimit}
    W_{1}( \xi; \ket{0,M})= \frac{1}{\left( 2 \pi \xi\right)^2} \int_{0}^{\infty}d\nu \frac{\nu}{e^{2 \pi \nu} - 1}. 
\end{align}
Let us discuss the interpretation given by the inertial observer for the excitation process. Since for the inertial observer the state of the field is the Minkowski vacuum, the detector makes transitions to excited states and quanta of the field show up. The source of energy that allows for these processes comes from the agent that accelerates the detector. Applying Eq.~\eqref{eq:bogoliubov1} to the Minkowski vacuum state, we get
\begin{equation}\label{eqq:barelation}
    b(\nu,\mathbf{q})\ket{0,M}=\int d^{3}\mathbf{k}\,V(\nu,\mathbf{q},\mathbf{k})a^{\dagger}(\mathbf{k})\ket{0,M}.
\end{equation}
The traditional way to interpret the above equation is to assert that counter-rotating processes $\sigma^{+}a^{\dagger}(\mathbf{k})$ ($\sigma^{+}=\ket{e}\bra{g}$ for usual two-level detectors) are responsible for the right-hand side contribution. These represent a virtual process since these occur for short time intervals $\Delta \tau$ obeying $|\Delta \omega| \Delta \tau < 1$, where $\Delta \omega$ is the associated energy gap, causing the detector to ``react" to the vacuum fluctuations~\cite{svaiter1992}. For a detector at rest in a rectilinear uniformly accelerated  moving frame, this process becomes a real one, which is commonly deemed as Unruh radiation. The same analysis performed by an accelerated observer can be realized by the inertial observer. The $F_1$ contribution is a process of excitation of the detector with emission.

On the other hand, in order to understand why inertial observers should perceive the existence of radiation emitted by a uniformly accelerated detector, one may also resort to the familiar result derived from the interaction of a quantized electromagnetic field with a classical source~\cite{Zuber05}. Indeed, in this case one verifies that classical current creates a coherent state from the vacuum. The probability corresponding to the emission of photons is given by a Poisson distribution. Even though such physical situations should be plainly distinguished, one may argue that in both cases the physical interpretation according to the inertial-observer perspective should follow along similar lines. In this regard, an interesting investigation is the study of the result of emission of quanta by a classical current for the case of a uniformly accelerated current coupled with a quantum field. This is a relevant subject and we will carefully examine this calculation in the near future.

\section{The Glauber detector in Kalnins-Miller coordinate system}\label{sec:kalnis}

The aim of this section is to discuss the behaviour of a detector in a frame that in the remote past is inertial and in the far future becomes uniformly accelerated. A coordinate system adapted to this frame was obtained in Refs. \cite{Boyer:1974am,boyer1976symmetry,kalnins1977lie}. The scalar field quantization was discussed by Costa, Svaiter and De Paola \cite{Costa:1987vv, Costa:1989, costa1989separable, svaiter1989,DePaola:1996xw}. For the four-dimensional spacetime  with the usual cartesian coordinates $x^{\mu} = (t, x,\mathbf{y})$ and the curvilinear coordinates $X^{\mu} = (\xi, \eta,\mathbf{y})$, we define the following mapping
\begin{align}
    t + x &=\frac{2}{a} \sinh{a(\xi+\eta)},
\end{align}
and
\begin{align}
     t - x &=-\frac{1}{a} e^{-a(\xi-\eta)},
\end{align}
for $-\infty< \eta < \infty$ and $-\infty< \xi < \infty$. This coordinate system $X^{\mu} = (\xi,\eta,\mathbf{y})$ is valid only for $t - x < 0$. Nevertheless, it is possible to extend it in order to cover the whole spacetime. The four-dimensional line element can be written using this curvilinear coordinates. We get 
\begin{equation}
    ds^2 = \left(e^{- 2 a \eta} + e^{2 a \xi} \right)\left( d \eta^2 - d \xi^2\right)-dy^{2}-dz^{2}.
\end{equation}
The next step is to discuss the proper acceleration of an observer travelling in the worldline $\xi, \mathbf{y} = \text{constant}$. We have that
\begin{equation}
    \alpha = a\left. \left(e^{- 2 a \eta} + e^{2 a \xi} \right)^{-3/2} e^{2 a \xi}\right|_{\xi = \bar{\xi}, \mathbf{y} = \bar{\mathbf{y}}}. 
\end{equation}
Therefore, in the remote past and in the far future we have, respectively 
\begin{equation}
    \begin{cases}
    \lim\limits_{\eta \rightarrow - \infty} \left.\alpha(\eta, \xi,\mathbf{y})\right|_{\xi = \bar{\xi}, \mathbf{y} = \bar{\mathbf{y}}} = 0,\\
    \lim\limits_{\eta \rightarrow  \infty} \left.\alpha(\eta, \xi,\mathbf{y})\right|_{\xi = \bar{\xi}, \mathbf{y} = \bar{\mathbf{y}}} = a e^{- a \bar{\xi}} = a_{\infty}.
    \end{cases}
\end{equation}
In these curvilinear coordinates the Klein-Gordon can be written as 
\begin{equation}
    \left[\frac{\partial^{2}}{\partial \eta^{2}}-\frac{\partial^{2}}{\partial \xi^{2}}+m_{0}^{2}\left(e^{-2 a \eta}+e^{2 a \xi}+\mathbf{q}^{2}\right)\right] \phi(\eta, \xi,\mathbf{y})=0.
\end{equation}
Let us define the variables $\zeta = a^{-1}e^{- a \eta}$ and $ \chi = a^{-1} e^{a \xi}$ for $\infty > \zeta > 0$ and $0< \chi< \infty$. Using the result obtained by Kalnins and Miller we write $\phi(\zeta,\chi, \mathbf{y})=F(\zeta)G(\chi)H(\mathbf{y})$. The separation of variables of the above equations yields 
\begin{equation}
    \left[\frac{d^{2}}{d \zeta^{2}}+\frac{1}{\zeta} \frac{d}{d \zeta}+m_{0}^{2}+\mathbf{q}^{2}+\frac{\lambda^{2}}{\zeta^{2}}\right] F(\zeta)=0,
\end{equation}
and
\begin{equation}
    \left[\frac{d^{2}}{d \chi^{2}}+\frac{1}{\chi} \frac{d}{d \chi}-m_{0}^{2}-\mathbf{q}^{2}+\frac{\lambda^{2}}{\chi^{2}}\right] G(\chi)=0.
\end{equation}
There are two complete orthonormal bases that can be used to expand the scalar field, $\{u_{\lambda}(\zeta,\chi,\mathbf{y}), u^{*}_{\lambda}(\zeta,\chi,\mathbf{y})\}$ and $\{v_{\lambda}(\zeta,\chi,\mathbf{y}), v^{*}_{\lambda}(\zeta,\chi,\mathbf{y})\}$ which are of the form
\begin{align}
    u_{\lambda}(\zeta,\chi,\mathbf{y})&=N_{\lambda} H_{i \lambda}^{(1)}\left(m \zeta\right) K_{i \lambda}(m \chi)e^{i\mathbf{q}\cdot\mathbf{y}},\\
    u_{\lambda}^{*}(\zeta,\chi,\mathbf{y})&=N_{\lambda} H_{-i \lambda}^{(2)}(m \zeta) K_{i \lambda}(m \chi)e^{-i\mathbf{q}\cdot\mathbf{y}},
\end{align}
where $m=\sqrt{m_{0}^{2}+\mathbf{q}^{2}}$ for
\begin{equation}
    N_{\lambda}= \frac{1}{4\pi^{3/2}}\left[\lambda\left(1-e^{-2 \pi \lambda}\right)\right]^{1 / 2},
\end{equation}
and
\begin{align}
    v_{\nu}(\zeta,\chi,\mathbf{y})&=\left(\frac{\nu}{4\pi^{3}}\right)^{1 / 2} J_{i \nu}(m \zeta) K_{i \nu}(m \chi)e^{i\mathbf{q}\cdot\mathbf{y}},\\
    v_{\nu}^{*}(\zeta,\chi,\mathbf{y})&=\left(\frac{\nu}{4\pi^{3}}\right)^{1 / 2} J_{-i \nu}(m \zeta) K_{i \nu}(m \chi)e^{-i\mathbf{q}\cdot\mathbf{y}},
\end{align}
where $H^{(1)}_{ \nu}(z)$, $H^{(2)}_{\nu}(z)$ are Bessel functions of third kind or Hankel's function and $J_{\nu}(z)$ is a Bessel function of first kind.  One can show that $u_{\lambda}(\zeta,\chi,\mathbf{y})$ are $u^{*}_{\lambda}(\zeta,\chi,\mathbf{y})$ are positive and negative frequency modes in the remote past and $v_{\nu}(\zeta,\chi,\mathbf{y})$ are $v^{*}_{\nu}(\zeta,\chi,\mathbf{y})$ are respectively positive and negative frequency modes in the far future \cite{disessa1974,Sommerfield:1974fa}. In the remote past, the field can be expanded as 
\begin{align}
    \phi(\zeta,\chi,\mathbf{y})=\int_{0}^{\infty}d\lambda\int d^{2}\mathbf{q}&\left[a_{\text{in}}(\lambda,\mathbf{q})u_{\lambda}(\zeta,\chi,\mathbf{y})\right. \nonumber\\ &+\left.a_{\text{in}}^{\dagger}(\lambda,\mathbf{q})u_{\lambda}^{*}(\zeta,\chi,\mathbf{y})\right], 
\end{align}
where $a_{\text{in}}(\lambda,\mathbf{q})$ and $a^{\dagger}_{\text{in}}(\lambda,\mathbf{q})$ are annihilation and creation operator for quanta of the field  in the remote past. It is now possible to define a vacuum state by 
\begin{equation}
    a_{\text{in}}(\lambda,\mathbf{q}) \ket{0, \text{in}} = 0, \ \forall \lambda,\mathbf{q}. 
\end{equation}
In the same way, the field expansion suitable for observers in the far future can be written as
\begin{align}
     \phi(\zeta,\chi,\mathbf{y})=\int_{0}^{\infty}d\nu\int d^{2}\mathbf{q}&\left[a_{\text{out}}(\nu)v_{\nu}(\zeta,\chi,\mathbf{y})\right. \nonumber \\
     &+\left. a_{\text{out}}^{\dagger}(\nu)v_{\nu}^{*}(\zeta,\chi,\mathbf{y})\right],
\end{align}
where $a_{\text{out}}(\nu,\mathbf{q})$ and $a^{\dagger}_{\text{out}}(\nu,\mathbf{q})$ are annihilation and creation operator for quanta of the field  in the far future. The vacuum state in this case is defined by
\begin{equation}
    a_{\text{out}}(\nu,\mathbf{q}) \ket{0, \text{out}} = 0, \ \forall \nu,\mathbf{q}. 
\end{equation}
The number of quanta associated with the $(\nu, \mathbf{q})$-out modes in the vacuum defined in the remote past is given by
\begin{equation}
    \bra{0,\text{in}}  a_{\text{out}}^{\dagger}(\nu,\mathbf{q}) a_{\text{out}}(\nu,\mathbf{q}) \ket{0,\text{in}} = \int d\lambda\int d^{2}\mathbf{q}' \vert \beta_{\nu\mathbf{q},\lambda\mathbf{q'}}\vert^{2},
\end{equation}
where the Bogoliubov coeficients are given by
\begin{equation}
    \vert \beta_{\nu\mathbf{q},\lambda\mathbf{q'}}\vert=\frac{1}{a\sqrt{e^{2\pi\nu}-1}}\delta(\nu-\lambda)\delta(\mathbf{q}-\mathbf{q}').
\end{equation}
This model presents a behaviour similar to the Hawking effect. If one considers the quantum version of a field theory without interactions the vacuum state constructed in the infinity past appears as a thermal state in the far future. Let us discuss the behaviour of a Glauber detector in this model. We define the quantity
\begin{equation}
    W(x;\ket{\Phi})= \bra{\Phi} :\varphi^{(-)}(x)\varphi^{(+)}(x):_{\text{out}} \ket{\Phi}.
\end{equation}
As discussed above, for a broadband detector this is proportional to the excitation rate. For $\ket{\Phi} = \ket{0, \text{out}}$ we get $W = 0$ as expected. We are interested in considering the case 
\begin{equation}
    W(x;\ket{0, \text{in}}) = \bra{0, \text{in}} :\varphi^{(-)}(x)\varphi^{(+)}(x):_{\text{out}} \ket{0, \text{in}}.
\end{equation}
A straightforward calculation gives us 
\begin{align}
    W(\zeta, \chi;\ket{0, \text{in}}) = &\frac{1}{4\pi^{3}} \int d^{2}\mathbf{q} \int_{0}^{\infty} d\nu\frac{\nu}{e^{2\pi \nu} - 1} \nonumber \\
    &J_{i\nu}(m \zeta)J_{-i\nu}(m \zeta)K_{i \nu}^{2}(m \chi).
\end{align}
For the case where $m_{0}^2 = 0$ and $\zeta \rightarrow 0$, we get
\begin{align}\label{eq:masslesslimit2}
   \lim\limits_{\zeta \rightarrow 0} W(\zeta, \chi;\ket{0, \text{in}})= \frac{1}{( 2 \pi \chi)^2} \int_{0}^{\infty}d\nu \frac{\nu}{e^{2 \pi \nu} - 1}. 
\end{align}
Hence the same interpretation given to $W_{1}$ and $W_{\beta}$ can be discussed in the Kalnins-Miller model. Previously we have obtained that the transition probability rate of an accelerated detector interacting with a thermal Rindler bath is equal to the transition probability rate of the accelerated detector coupled to the field in the Minkowski vacuum. In a similar fashion, we observe that, for the massless field, the transition probability rate of the detector in the far future is equivalent to the transition probability rate of the detector at rest in a non-inertial reference frame interacting with the field prepared in an usual thermal state. 

One can ask how to generate the Rindler vacuum state. It is known that accelerated mirrors can be used for such a purpose~\cite{PhysRevLett.121.071301}. The Kalnins-Miller model can be employed to investigate the accelerated mirror problem imposing a Dirichlet boundary conditions in one worldline $\chi = \text{constant}$ in a two dimensional toy model. See, for example, the approach presented in Ref.~\cite{fulling.davies.1976}. A natural continuation is to discuss the flux of radiation from this mirror.


\section{Conclusions}\label{sec:conclusions}

In this paper, we have proposed to study the Unruh-Davies effect using the Glauber theory of photodetection. There are some simplifications in our discussion such as the use of point-like interactions (fields are operator-valued distributions) and the absence of a switching function. We believe that it is possible to understand the effect even with such simplifications. We discussed the process of measuring Rindler field quanta by a uniformly accelerated detector interacting with a massive scalar field. Using the Glauber theory of photodetection, the rate of excitation of an accelerated detector interacting with the scalar field was evaluated.  We rederive the result that a uniformly accelerated detector coupled to a field in the Minkowski vacuum state has the same transition rate of an accelerated detector interacting with the field in a Rindler thermal one at some temperature $\beta^{-1}$.

Our results show that an inertial interpretation of the radiative processes associated with observers at rest in the frame of the uniformly accelerated detector is available in the Glauber formalism. In other words, we showed how the inertial observer interprets the excitation of the accelerated detector. The fundamental difference between the Unruh-DeWitt detector and the Glauber detector is the contribution coming from $F_{2}$.  In any case we have given above detailed arguments that showed that such a term does not actually contribute to the excitation rate for large observation time intervals.

Also using the Glauber theory of photodetection we have discussed a  model with unitary non-equivalent vacuum states in the remote past in the far future. This detector presents a behaviour similar to the detector interacting with a field in a scenario of gravitational collapse. For the massless case, we obtained that the transition probability rate of the detector in the far future is equivalent to the transition probability rate of the detector at rest in a non-inertial reference frame interacting with the field prepared in an usual thermal state. In this scenario, the introduction of a boundary condition allows one to investigate how to generate the Fulling vacuum. 

An interesting application of our analysis is the discussion of the equivalence principle in the quantum domain \cite{Fulling_2018, ben2019unruh}. One must show that the following physical situations are equivalent: The excitation rate of a detector at rest in a local inertial frame coupled to a field in the Boulware vacuum must be equal to the excitation rate of a detector travelling in a inertial world line in a Minkoswki spacetime interacting with a field in the Fulling vacuum state \cite{PhysRevD.11.1404}. Since Rindler operators span only a sub-algebra of the equal-time field algebra, it should be clear that it is not straightforward to obtain the desired result. We hope to fully explore this issue in a future work.

\section*{Acknowledgments}
We would like to thank B. F. Svaiter, Alexis Hernandez and Gast\~{a}o Krein for useful discussions. We also thank the anonymous referees for numerous relevant suggestions. This work was partially supported by Conselho Nacional de Desenvolvimento Cient\'{\i}fico e Tecnol\'{o}gico -- CNPq, under grants 309982/2018-9 (C.A.D.Z.), 303436/2015-8 (N.F.S.) and 310291/2018-6 (G.M.), and Funda\c{c}\~ao Carlos Chagas Filho de Amparo \`a Pesquisa do Estado do Rio de Janeiro -- FAPERJ under grant E-26/202.725/2018 (G.M.).

\end{document}